\documentclass[a4paper,11pt]{article}
\usepackage[margin=2.5cm]{geometry}
\usepackage[labelfont=bf]{caption}

\usepackage[labelfont=bf,font=small, justification=raggedright]{caption}
\usepackage{pifont}
\usepackage{cancel}
\usepackage{hyperref}
\usepackage{graphicx}
\usepackage[utf8]{inputenc}
\usepackage{amsmath}
\usepackage[normalem]{ulem}
\usepackage{verbatim}
\usepackage{amssymb}
\usepackage{listings}
\usepackage{hyperref}
\usepackage{mathtools}
\usepackage{color}
\usepackage{xspace}
\usepackage{lipsum}
\usepackage{tikz-cd}
\usepackage{ifthen}
\usepackage{float}
\usepackage{esvect}
\usepackage{dcolumn}
\usepackage[all]{nowidow}

\newcommand{\prog}[1]{Alg.~\ref{alg:#1} (\sub{#1})}

\newcommand{\progg}[1]{Algorithm~\ref{alg:#1} (\sub{#1})}

\newcommand{\progn}[1]{Alg.~\ref{alg:#1}}

\newcommand{\pMet}{\PCAL^{\text{Met} }}
\newcommand{\pFact}{\PCAL^{\text{fact}}}
\newcommand{\pFactp}{\PCAL^{\text{fact}+}}
\newcommand{\pFactm}{\PCAL^{\text{fact}-}}

\newcommand{\equidist}{equilibrium pair distance $b$\xspace}

\newcommand{\bcrit}{b_\text{crit}}

\newcommand{\xmean}{\overline{x}}
\newcommand{\PROCEDURE}[1]{\textbf{procedure}\ \sub{#1}}

\newcommand{\BRACE}[1]{
\;\;\;  \left\{\begin{array}{l}#1\end{array} \right.}
\newcommand{\IS}[2]{#1 \leftarrow #2}

\newcommand{\FOR}[1]{\textbf{for}\ #1 \textbf{: }}

\newcommand{\ENDPROCEDURE}{\text{------} \\ \vspace{-0.8cm}}

\newcommand{\IF}[1]{\textbf{if } #1 \textbf{: }}

\newcommand{\AND}{\textbf{and}}

\newcommand{\ELSE}{\textbf{else: }}

\newcommand{\OUTPUT}[1]{\textbf{output}\ #1}

\newcommand{\INPUT}[1]{\textbf{input}\ #1}
\newcommand{\COMMENT}[1]{\text{\footnotesize (#1)}}

\newcommand{\SET}[1]{\{#1\}}

\newcommand{\sub}[1]{\texttt{#1}}
\newcommand{\eq}[1]{eq.~(\ref{#1})}

\newcommand{\eqtwo}[2]{eqs.~(\ref{#1}) and~(\ref{#2})}
\newcommand{\eqfromto}[2]{eq.~(\ref{#1}) to eq.~(\ref{#2})}

\newcommand{\fig}[1]{Fig.~\ref{#1}}
\newcommand{\quot}[1]{``#1''}
\newcommand{\tab}[1]{Table~\ref{#1}} 
 
\newcommand{\app}[1]{Appendix~\ref{#1}}
 
\newcommand{\SECT}[1]{Section~\ref{#1}}

\newcommand{\vs}{\textrm{vs.}}

\newcommand{\ACAL}{\mathcal{A}}  
\newcommand{\OCAL}{\mathcal{O}}  
\newcommand{\PCAL}{\mathcal{P}}  
\newcommand{\expa}[1]{\mathrm{e}^{#1}}   
\newcommand{\expb}[1]{\exp \glb #1 \grb} 
\newcommand{\expc}[1]{\exp \glc #1 \grc} 
\newcommand{\ran}{\texttt{ran}}
\newcommand{\ranb}[2][]{\ran_{#1} \! \glb #2 \grb}  

\newcommand{\loga}[2][]{\log^{#1}\! \gla #2 \gra}  
\newcommand{\logc}[2][]{\log^{#1} \glc #2 \grc}  

\newcommand{\maxc}[2][]{\max^{#1} \glc #2 \grc}  

\newcommand{\minb}[2][]{\min^{#1} \glb #2 \grb}  
\newcommand{\minc}[2][]{\min^{#1} \glc #2 \grc}  
\newcommand{\mind}[2][]{\min^{#1} \gld #2 \grd}  

\newcommand{\gla}{\,}  
\newcommand{\gra}{}  
\newcommand{\glb}{\left(}  
\newcommand{\grb}{\right)}  
\newcommand{\glc}{\left[}  
\newcommand{\grc}{\right]}  
\newcommand{\gld}{\left\{}  
\newcommand{\grd}{\right\}}  
\newcommand{\gle}{\left|}  
\newcommand{\gre}{\right|}  

\newcommand{\const}{\text{const}}

\newcommand{\TO}{,\ldots,}

\newcommand{\ptilde}{\tilde{p}}

\newcommand{\xtilde}{\tilde{x}}
\newcommand{\ytilde}{\tilde{y}}

\newcommand{\Etilde}{\tilde{E}}

\newcommand{\mean}[1]{\left\langle #1 \right\rangle}
\newcommand{\half}{\frac{1}{2}}

\newcommand\bigOb[1]{\ensuremath{\OCAL\glb #1 \grb}}

\newcommand\diff[1]{\mathrm{d}#1}
\newcommand{\fracb}[2]{\frac{#1}{#2}}

\newcommand\subcap[1]{{(#1):}}
\newcommand{\REF}[2][]{
	\ifthenelse{\equal {#1} {}}{Ref.~\cite{#2}}{Ref.~\cite[#1]{#2}}}
\renewcommand{\emph}[1]{\textit{#1}}
\newcommand{\pacc}{p_{\text{acc}}}
\newcommand{\vpointer}{v_{\vv{x}}}
\newcommand{\vpmean}{\mean{v_{\vv{x}}}}

\newcolumntype{L}{D{.}{.}{2,5}}

\hypersetup{
    colorlinks=true,
    linkcolor=blue,
    filecolor=magenta,      
    urlcolor=cyan,
    pdftitle={A synopsis of HMC and LiftedMCMC}
}

\title{ Hamiltonian Monte Carlo \vs\ event-chain Monte Carlo: \\
an appraisal of sampling strategies beyond the diffusive regime}

\author{Werner Krauth\\
Laboratoire de Physique de l’Ecole normale
supérieure, \\
ENS, Université PSL, CNRS, Sorbonne Université, Université Paris Cité,
 Paris, France \\
Rudolf Peierls Centre for Theoretical Physics, Clarendon
 Laboratory, \\
 University of Oxford, Oxford OX1 3PU, UK \\
Simons Center for Computational Physical Chemistry, \\
New York University, New York (NY), USA}

\begin{document}
\newfloat{algorithm}{ht}{loa}
\floatname{algorithm}{Algorithm }
\setcounter{algorithm}{0}

\maketitle

\begin{abstract}
We discuss Hamiltonian Monte Carlo (HMC) and event-chain Monte Carlo (ECMC) for
the one-dimensional chain of  particles with harmonic interactions and benchmark
them against local reversible Metropolis algorithms. While HMC
achieves considerable speedup with respect to local reversible Monte Carlo
algorithms, its autocorrelation functions of global observables such as the
structure factor have slower scaling with system size than for ECMC, a lifted
non-reversible Markov chain. This can be traced to the dependence of ECMC on a
parameter of the harmonic energy, the equilibrium distance, which drops out when
energy differences or gradients are evaluated.   We review the recent literature
and provide pseudocodes and Python programs. We finally discuss related models
and generalizations beyond one-dimensional particle systems.
\end{abstract}
\tableofcontents
    
\section{Introduction}

Markov-chain Monte Carlo (MCMC) permeates all fields of science, from physics to
statistics and to the social disciplines. Reversible Markov chains, the great
majority of MCMC methods, map the sampling of probability distributions onto the
simulation of fictitious physical systems in thermal equilibrium. In physics,
thermal equilibrium is characterized by time-reversal invariance and the
detailed-balance condition. It thus comes as no surprise that reversible Markov
chains, such as the famous Metropolis~\cite{Metropolis1953} and heat-bath
algorithms~\cite{Glauber1963,Creutz1980HeatBath,Geman1984}, all satisfy detailed
balance. In physics, again, thermal equilibrium is characterized by diffusive,
local, motion of particles. This translates into the slow mixing and
relaxation dynamics of local reversible Markov chains, and it affects
most Markov chains used in practice.

In this paper, we appraise diametrically opposite strategies to overcome the
slow diffusive motion of local, reversible MCMC methods. One
is Hamiltonian Monte Carlo (HMC), a \emph{non-local} yet reversible Markov
chain, and the other is event-chain Monte Carlo (ECMC), a class of lifted
Markov chains, which are local yet \emph{non-reversible}.

HMC~\cite{DuKePeRo1987} emulates classical physical systems which, even when
they
are out of equilibrium, satisfy Newton's (or Hamilton's) equations. In addition
to their positions (the sample-space variables of the fictitious physical
systems), particle systems have momenta. While the positions change in time as
dictated by the momenta, momenta change with the gradient of a potential.
Approaching equilibrium then corresponds to a gradient descent, which is
generally assumed to be fast because it corresponds to a physical dynamics. In a
nutshell, HMC introduces fictitious momenta and solves discretized Newton's
equation to move from one configuration $x$ to a distant configuration $x'$. The
time-discretization errors are eliminated through rejections  which enforce
detailed balance.

ECMC~\cite{Bernard2009,Michel2014JCP} introduces additional
\quot{lifting} variables which are more
general than momenta. As the name indicates, ECMC  is an event-driven
implementation of a continuous-time (thus local) Markov process. It thus need
not discretize time, and exactly satisfies global balance, a continuity
condition which replaces detailed balance in the non-reversible case. The ECMC
dynamics is neither diffusive nor gradient-based. In fact, it evaluates no
global potential, potential differences or gradients, and it crucially depends
on a parameter (a \quot{factor field}~\cite{Lei2019} or
\quot{pullback}~\cite{Essler2024}), which leaves the gradient invariant. For a
many-particle harmonic chain on a one-dimensional ring that we use as a
benchmark, ECMC has better scaling than HMC.

In \SECT{sec:HarmonicThermodynamics} of this paper, we review the harmonic
chain~\cite{Lei2019}, compute its thermodynamics, and present
basic sampling algorithms for it, on the one hand direct-sampling methods, and
on the other the aforementioned local reversible Markov chains featuring slow
diffusive dynamics. In \SECT{sec:HMC}, we implement an HMC algorithm for the
harmonic chain, following \REF{Neal2011}, and discuss the algorithm's stability
and its relation to the molecular-dynamics method. In \SECT{sec:ECMC}, we
implement several variants of ECMC for the harmonic chain, following
\REF{Lei2019}. We discuss the essential link~\cite{Michel2014JCP} between the
behavior of the algorithm and the system pressure and outline how the algorithm
has been generalized to more complex sampling problems. In
\SECT{sec:ComparisonsBenchmarks}, we juxtapose the different sampling strategies
and show how they effectively go beyond the diffusive regime and improve the
scaling of autocorrelation times of global observables, in our case the
structure factor. In the concluding \SECT{sec:Conclusion} we discuss the generic
nature of the harmonic-chain dynamics both for HMC and ECMC. We also discuss the
relation of ECMC for the harmonic chain with a lattice model, the \quot{lifted}
TASEP (totally asymmetric simple exclusion model), whose dynamics can be
analyzed exactly using the Bethe ansatz~\cite{Essler2024}. The main text of
this paper contains a
number of pseudocode algorithms.   The appendices provide mathematical details
(\app{app:Math}), as well as further pseudocode algorithms and pointers towards
an open-source repository containing all the  Python programs discussed in this
paper (\app{app:Algorithms}).

\section{Harmonic chain: thermodynamics and basic sampling}
\label{sec:HarmonicThermodynamics}

The harmonic chain describes particles on a continuous interval with periodic
boundary conditions in space and in the indices. Each particle interacts with
the two particles with neighboring indices. The interaction is harmonic with an
equilibrium pair distance that drops out for the Metropolis algorithm and for
HMC but is all important within ECMC. The harmonic chain is equivalent to the
time-discretized path integral of a free quantum particle~\cite{SMAC}.

In \SECT{subsec:DefinitionsThermodynamics}, we define the harmonic-chain model
and exactly compute its thermodynamics, in particular the pressure, whose
vanishing defines a special point for ECMC. We further discuss, in
\SECT{subsec:Levy}, direct-sampling algorithms for the harmonic chain and,  in
\SECT{subsec:GibbsMetropolis}, examples of local reversible Markov chains
featuring the slow dynamics that HMC and ECMC overcome in different ways. The
Metropolis algorithm is the classic sampling approach, whereas the factorized
Metropolis algorithm is the starting point for the nonreversible ECMC.

\subsection{Model definition, thermodynamics}
\label{subsec:DefinitionsThermodynamics}
The harmonic chain consists of  $N$ particles $k \in \SET{0 \TO N-1}$ on a ring
of length $L$ with periodic boundary conditions. For  finite $L$, a
convenient description is in terms of one periodic coordinate
\begin{gather}
 0 \le x_0 < L, \\
 \intertext{and of elongations $x_k - x_{k-1}$ with }
 -\infty < x_1 \TO x_{N-1} < \infty. \\
\intertext{Periodic boundary conditions are implemented through}
 x_N = x_0 + L .
 \label{equ:PeriodicN}
\end{gather}
Each configuration $x = \SET{x_0 \TO x_{N-1} }$ (with $x_N$ from
\eq{equ:PeriodicN} understood) has a harmonic potential
\begin{align}
 U(x, b, L)  &= \half \sum_{k=1}^{N} (x_k - x_{k-1} -b)^2\quad
\label{equ:UofxB}\\
           &= \half \sum_{k=1}^{N} (x_k - x_{k-1})^2 - b \sum_{k=1}^N (x_k -
           x_{k-1}) + \half N b^2 \label{equ:UofxFactorFields}\\
           &= \half \sum_{k=1}^{N} (x_k - x_{k-1})^2 - b L + \half N b^2,
\label{equ:UofxIndependent}
\end{align}
with the \equidist. In \eq{equ:UofxIndependent}, the terms in $b$ are the same
for
all particle positions, so that algorithms that rely on potential
differences and gradients, such as HMC, are insensitive to them. They play
nevertheless a crucial role for ECMC dynamics~\cite{Lei2019}.

The Boltzmann weight of each configuration is given by
\begin{equation}
 \pi(x, b, L) = \expc{- \beta U(x, b, L)},
\label{equ:Boltzmann}
\end{equation}
with the partition function
\begin{align}
 Z(N, b, L) &= \int_{0}^{L} \diff x_0
 \int_{-\infty}^{\infty} \diff x_1 \dots
 \int_{-\infty}^{\infty} \diff x_{N-1} \pi(x,b,L) \\
&= L \expc{\beta b L - \half \beta b^2 N - \half \beta L^2 N }
 \frac{2^{(N-1)/2}}{\sqrt{N}}
\frac{\pi^{(N-1)/2}}{\beta^{(N-1)/2}}
\label{equ:PartitionFunction}
\end{align}
(see \app{app:Math} for a derivation). The pressure is
\begin{equation}
 P|_{\beta = 1} = \partial \log Z / \partial L|_{\beta = 1}  =   \frac{1}{L} + b
 - \frac{L}{N}.
\label{equ:FabulousFormula}
\end{equation}
For $b =  \bcrit =  L/N - 1/L$, the pressure vanishes. This defines a special
point for
ECMC.
The mean energy is
\begin{equation}
\mean{U} |_{\beta = 1} = -\partial \log Z / \partial \beta |_{\beta = 1} =  -bL
+ \half b^2 N + \half \frac{L^2}{N} + \frac{(N-1)}{2}.
\label{equ:MeanEnergy}
\end{equation}
Its $b$-independent part (the last two terms on the right-hand side)
will be used for testing our computer implementations.

The gradient of the energy,
\begin{equation}
\nabla_k  U(x, b, L)  =  \begin{cases}
2 x_0 - \glb x_{1} + x_{N-1} - L \grb & \text{for $k=0$} \\
2 x_k - \glb x_{k+1} + x_{k-1} \grb & \text{for $0< k< N-1$}\\
2 x_{N-1} - \glb x_{0} + L  + x_{N-2} \grb & \text{for $k = N-1$}
                       \end{cases},
\label{equ:GradientHarmonic}
\end{equation}
is independent of $b$. The harmonic chain is defined in terms of distinguishable
particles, which allows to define elongations $x_k - x_{k-1}$ between $\pm
\infty$). Throughout this paper, we only consider observables which
are invariant under a
uniform translation of the positions. This has been formalized as a difference
between a \quot{configuration} and a \quot{state} of a Markov
chain~\cite{RandallWinklerCircle2005}.

\subsection{Direct sampling: Lévy construction (Gaussian bridge)}
\label{subsec:Levy}
The Boltzmann distribution $\pi(x,b,L)$ of \eq{equ:Boltzmann} can be sampled
directly, without resorting to Markov chains. One direct-sampling algorithm, the
Lévy construction~\cite{Levy1940} (also known as the \quot{Gaussian bridge}), is
a staple in fields from path-integral Monte Carlo~\cite{Pollock1987,Krauth1996}
to financial mathematics~\cite{Glasserman2003}. The position $x_0$ is sampled as
a uniform random variables between $0$ and $L$. Then, as $x_N$ is known through
\eq{equ:PeriodicN}, one can iteratively sample a Gaussian $x_1$ (conditioned on
$x_0$ and $x_N$) then another Gaussian $x_2$ (conditioned on $x_1$ and $x_N$)
and so on until $x_{N-1}$ (see \prog{levy-harmonic}). The algorithm exists in
many variants, including in Fourier space~\cite[Chap. 3]{SMAC}. All these
algorithms are naturally independent of the \equidist, reinforcing the naive
conception that $b$ is irrelevant for sampling. We use the Lévy construction to
initialize our Markov chains in equilibrium.

\begin{algorithm}
    \newcommand{\algo}{levy-harmonic}
    \captionsetup{margin=0pt,justification=raggedright}
    \begin{center}
        $\begin{array}{ll}
           & \PROCEDURE{\algo}\\
           & \IS{x_0}{\ranb{0,L}} \\
           & \IS{x_N}{x_0 + L} \\
           & \FOR{k = 1 \TO N-1}\\
           & \BRACE{
            \IS{\xmean}{[(N-k) x_{k-1} + x_N]/(N-k+1)}\\
            \IS{\sigma}{\sqrt{1/[1 + 1/(N-k)]} }\\
            \IS{x_k}{\sub{gauss}(\xmean, \sigma)  }\\
           }\\
           & \OUTPUT{\SET{x_0 \TO x_{N-1}}}\
\COMMENT{configuration, sample of \eq{equ:Boltzmann}}\\
           & \ENDPROCEDURE\
        \end{array}$
    \end{center}
    \caption{\sub{\algo}. Lévy construction (Gaussian bridge) providing
    a direct-sampling algorithm for the harmonic chain. This algorithm
    initializes all the Markov chains in this paper.}
\label{alg:\algo}
\end{algorithm}

\subsection{Metropolis and factorized Metropolis algorithms}
\label{subsec:GibbsMetropolis}

Any Markov chain that is irreducible and aperiodic converges
to the distribution $\pi$ if it satisfies the global-balance
condition~\cite{Levin2008}
\begin{equation}
\pi(x') = \sum_{x}\pi(x) P(x, x').
\label{equ:GlobalBalance}
\end{equation}
Here, the transition matrix $P(x, x')$ is the probability to move from a given
point $x$ to $x'$ in one time step, from time $t$ to $t+1$, conditioned on the
Markov chain being at position $x$  at time
$t$. A stricter condition for convergence is the
detailed-balance condition
\begin{equation}
 \pi(x) P(x, x') = \pi(x') P(x', x),
 \label{equ:DetailedBalance}
\end{equation}
which implies \eq{equ:GlobalBalance} because of the conservation
of probabilities $\sum_{x'}P(x, x') = 1$. Commonly, the transition matrix $P(x,
x')$
is written as a product
\begin{equation}
P(x,x') = \ACAL(x,x') \PCAL(x, x').
\end{equation}
Here, the \emph{a priori} probability $\ACAL(x, x')$ \emph{proposes} the move $x
\to x'$, and the filter $\PCAL(x, x')$
\emph{accepts or rejects} it. The Metropolis algorithm uses a symmetric
$\ACAL$,
\begin{equation}
 \ACAL(x, x') = \ACAL(x', x),
 \label{equ:APrioriSymmetric}
\end{equation}
with a specific choice of symmetric function:
\begin{equation}
 \pi(x) \pMet(x, x') = \pi(x') \pMet(x',  x) = \underbrace{\minc{\pi(x),
\pi(x')}}_{\text{symmetric in $x$ and $x'$}}.
\end{equation}
As the right-hand side of this equation is symmetric in $x$ and $x'$, so must
be the rest of the equation, so that detailed balance is satisfied.
This leads to the famous Metropolis filter
\begin{equation}
 \pMet(x, x') =
 \minc{1, \frac{\pi(x')}{\pi(x)}} =
 \mind{1, \expa{-\beta [U(x') - U(x)]}}.
\end{equation}
In the harmonic chain, for a move  $x_k \to x_k '$ of a single particle $k$, the
change of the potential $U$, written as in \eq{equ:UofxB}, impacts two terms,
namely
\begin{align}
U_k^+ = \half (x^+ - x_k - b )^2;\quad  & U_k^- = \half (x_k - x^- -b)^2, \\
\intertext{in the initial configuration containing $x_k$ and two terms, namely}
U_k{'}^{+} = \half (x^+ - x_k' - b )^2;\quad  & U_k^- = \half (x_k' - x^-
-b)^2,
\end{align}
for the final configuration containing $x_k'$
(up to boundary conditions in space and in particle index, $x^+ = x_{k+1}
$ and $x^- = x_{k-1} $). Writing $U_k = U_k^+ + U_k^-$ (and $U_k'$ for the
potentials of the new position), the Metropolis filter becomes
\begin{align}
 \pMet(x_k, x_k') =
& \mind{1, \expc{-\beta (U_k' - U_k)}} = \\
& \mind{1,
\expc{-\beta ({U^+_k}' - U^+_k)}
\expc{-\beta ({U^-_k}' - U^-_k)}
}
\label{equ:MetropolisFilter2}
 .
\end{align}
This is implemented for $\beta=1$ in \prog{metropolis-harmonic}, together
with a consistent treatment of periodic boundary conditions in position and
particle index. The algorithm is independent of $b$, because of
\eqtwo{equ:UofxB}{equ:UofxIndependent}.

\begin{algorithm}
    \newcommand{\algo}{metropolis-harmonic}
    \captionsetup{margin=0pt,justification=raggedright}
    \begin{center}
        $\begin{array}{ll}
           & \PROCEDURE{\algo}\\
           & \INPUT{\SET{x_0 \TO x_{N-1}}, t}\
            \COMMENT{configuration, time}\\
           & \IS{k}{\sub{choice}(\SET{0 \TO N-1})};\
           \IS{\Delta_x}{\ranb{-\delta, \delta}}\\
           & \IS{k^+}{\sub{mod}(k+1,N)},\
             \IS{k^-}{\sub{mod}(k-1,N)} \\
           & \IS{x^+}{x_{k^+}},\ \IF{k = N-1}{\IS{x^+}{x^+ + L}} \\
           & \IS{x^-}{x_{k^-}},\ \IF{k = 0}{\IS{x^-}{x^- - L}} \\
           & \IS{x_k'}{x_k + \Delta_x}\\
           & \IS{U_k}{\half \glc (x^+ - x_k)^2 + (x_k - x^-) ^2 \grc}\
\COMMENT{no need to incorporate $b$ (see
\eqtwo{equ:UofxB}{equ:UofxIndependent}}\\
           & \IS{U'_k}{\half \glc (x^+ - x'_k)^2 + (x'_k - x^-) ^2 \grc}\\
           & \IS{\Upsilon}{\ranb{0,1}}\\
           & \IF{\Upsilon < \expc{-(U_k'- U_k)}}\ \IS{x_k}{x'_k}\
          \COMMENT{equivalently: $\IF{\Upsilon < \pMet_k(x_k \to
x'_k)} ...$}
           \\
           & \OUTPUT{\SET{x_0 \TO x_{N-1}}, t+1}\
\COMMENT{configuration, time}\\
           & \ENDPROCEDURE\
        \end{array}$
    \end{center}
    \caption{\sub{\algo}. Metropolis Monte Carlo algorithm for the harmonic
chain, featuring local moves of single particles $k$ on a scale $\delta$.
The algorithm is independent of $b$.}
\label{alg:\algo}
\end{algorithm}

The \emph{factorized} Metropolis algorithm~\cite{Michel2014JCP} inherits the
symmetric \emph{a priori} probability from \prog{metropolis-harmonic}, but uses
a filter which is the product of two factor filters
\begin{align}
 \pFact(x_k, x'_k) =
 & \mind{1,
\expc{-\beta (U^{\prime+}_k - U^+_k)}}
\mind{1, \expc{-\beta (U^{\prime -}_k - U^-_k)}
} =  \\
 & \overbrace{
 \minc{1, \frac
 { \expa{-\frac{\beta}{2}
 (x^+ - x'_k - b)^2 }}
 { \expa{-\frac{\beta}{2}
 (x^+ - x_k - b)^2 }} }
 }^{\pFactp(x_k, x'_k)} \
 \overbrace{
 \minc{1, \frac
 { \expa{-\frac{\beta}{2}
 (x'_k - x^- - b)^2 }}
 { \expa{-\frac{\beta}{2}
 (x_k - x^-  - b)^2 }} }
 }^{\pFactm(x_k, x'_k)}.
 \label{equ:FactorizedFilter}
\end{align}
This filter may of course be  implemented by comparing a random
number $\Upsilon$ to $\PCAL^{\text{fact}}(x_k,  x'_k)$, but its true
potential stems from rather comparing \emph{two} independent random numbers to
two factor filters, namely $\Upsilon^+$  to $\PCAL^{\text{fact} +}(x_k,
x'_k)$ and $\Upsilon^-$  to $\PCAL^{\text{fact} -}(x_k, x'_k)$. The move is
accepted by consensus (see \REF{Tartero2024} for a discussion and
\prog{metropolis-factorized-harmonic}). The explicit $b$
dependence will later allow the non-reversible ECMC to improve the
scaling of
relaxation times. Run reversibly, as  in
\progn{metropolis-factorized-harmonic}, it does not have this advantage. In
fact,  the factorized variant is slower than the traditional Metropolis
algorithm simply because $\pFact(x_k,  x'_k) \le \pMet(x_k,  x'_k)$. It
thus requires somewhat smaller values of $\delta$ than
\progn{metropolis-harmonic} in order to keep up a sizeable acceptance rate.

\begin{algorithm}
    \newcommand{\algo}{metropolis-factorized-harmonic}
    \captionsetup{margin=0pt,justification=raggedright}
    \begin{center}
        $\begin{array}{ll}
           & \PROCEDURE{\algo}\\
           & \INPUT{\SET{x_0 \TO x_{N-1}}, t}\
            \COMMENT{configuration, time}\\
           & \dots  \COMMENT{$k,\Delta_x, k^+,k^-, x^+ , x^-, x_k' $ as in
\progn{metropolis-harmonic}}\\
           & \IS{U_k^+}{\half  (x^+ - x_k - b)^2};\
            \IS{U_k^-}{\half  (x_k - x^- - b)^2}\ \COMMENT{$b$ dependence not
to be dropped}  \\
           & \IS{U_k^{\prime +} }{\half  (x^+ - x_k' - b)^2};\
            \IS{U_k^{\prime -}}{\half  (x_k' - x^- - b)^2}\ \COMMENT{$b$
dependence
not to be dropped}  \\
           & \IS{\Upsilon^+}{\ranb{0,1}};\  \IS{\Upsilon^-}{\ranb{0,1}}\\
         * & \IF{\Upsilon^+ < \expc{-(U^{\prime +}_k- U^+_k)}\ \AND\
           \Upsilon^- < \expc{-(U^{\prime -}_k - U^{-}_k)}}
           \IS{x_k}{x'_k} \\
           & \OUTPUT{\SET{x_0 \TO x_{N-1}}, t+1}\
\COMMENT{configuration, time}\\
           & \ENDPROCEDURE\
        \end{array}$
    \end{center}
    \caption{\sub{\algo}. Factorized Metropolis algorithm for the harmonic
chain. The line * implements \eq{equ:FactorizedFilter}. The
algorithm depends on the parameter $b$.}
\label{alg:\algo}
\end{algorithm}

The dynamics of the factorized Metropolis algorithm explicitly depends on the
choice of factors. As a simple alternative factorization, we may write the
energy as in \eq{equ:UofxFactorFields} with four terms that change with a move
of  particle $k$:
\begin{align}
U_k^+ &= \half (x^+ - x_k)^2; & U_k^- = \half (x_k - x^-)^2; \ & U_f^+
= -b(x^+ - x_k);\ U_f^- = -b(x_k - x^-)
\label{equ:FourTermsInitial} \\
U_k^{\prime+} &= \half (x^+ - x_k')^2;\  & U_k^{\prime-} = \half (x_k' -
x^-)^2; \ & U_f^{\prime+}
= -b(x^+ - x_k');\ U_f^{\prime -} = -b(x_k' - x^-)
\label{equ:FourTermsFinal},
\end{align}
which then gives rise to four factor filters (see
\prog{metropolis-fourfactor-harmonic} in
\app{app:Algorithms} for an implementation). The term $b$ is referred to as a
\quot{factor field}~\cite{Lei2019}. It is related to the \quot{pullback} of the
lifted TASEP~\cite{Essler2024} and its generalizations~\cite{Essler2024b}, and
can be set up for general potentials $U(x)$.

\section{Hamiltonian Monte Carlo for the harmonic chain}
\label{sec:HMC}

\quot{Hamiltonian} Monte Carlo~\cite{DuKePeRo1987} (originally termed
\quot{hybrid}
Monte Carlo) implements a reversible Markov chain with a symmetric proposal
probability, just as the Metropolis algorithm. It proposes large moves  $x \to
x'$, but ensures that the acceptance probability remains high and thus
improves on the reversible Markov chains of \progn{metropolis-harmonic} and
\progn{metropolis-factorized-harmonic}. However, computing $x'$ given $x$ is
is time-consuming, and the computational complexity for
proposing the move $x
\to x'$ is determined by an instability which becomes more pronounced for
bigger sizes.

In \SECT{subsec:HMCAlgorithm}, we implement HMC for the harmonic chain and
sketch some of its basic properties. In \SECT{subsec:HMCStability}, we discuss
the optimal choice of parameters in order to propose and accept
non-local moves $x \to x'$ most efficiently. In
\SECT{subsec:HMCBasicRequirements}, we compare HMC to the
molecular-dynamics method.

\subsection{Implementation of HMC}
\label{subsec:HMCAlgorithm}
HMC associates the position $x$ at a given time $t$ with a set of
Gaussian
momenta $p = \SET{p_1 \TO p_N}$. Together,
they construct a probability distribution
\begin{equation}
 \pi(x, p) = \expc{-\beta U(x)} \expb{ -\beta p^2/2}.
 \label{equ:UandK}
\end{equation}
The energy $U$ thus becomes the potential energy of a dynamical system with
total energy $U + K$, where the kinetic energy is $K = \half p^2$.
An ideal Hamiltonian time evolution
moves in a time interval $\tau$ from $(x, p)$ to $(x', p')$. In the special
case of the hard-sphere
model, this dynamics samples the Boltzmann distribution
rigorously~\cite{Sinai1970,Simanyi2003}.
In more general problems, the (ideal) Hamiltonian time evolution
conserves the total energy $E = U + K$
but may not visit all configurations $(x,p)$ of energy $E$. In addition,
the potential energy $U$ satisfies $ U \le E$. With $E$ fixed, $U$ is also
bounded. It is thus evident that the momenta
$p$ must be resampled from \eq{equ:UandK}. In the aforementioned
hard-sphere models and variants, ideal Hamiltonian time evolution can be
implemented on the computer. This is the field of event-driven molecular
dynamics~\cite{Alder1957,Alder1962}, which applies to models with piecewise
constant potentials.
In the general case, however, the evolution
equations must be time-stepped, with
a time-discretization interval $\epsilon$. Symplectic time-stepping algorithms
do generally not conserve the energy $E$, so that the move $(x, p) \to
(\xtilde, \ptilde)$ generates an energy $\Etilde$. However, the algorithms used
in HMC are time-reversal invariant, and the Metropolis filter
establishes detailed balance with respect to the true Boltzmann distribution,
without any approximation or time-stepping error. The configuration $(x', p') $
equals $(\xtilde, \ptilde)$ if the move is accepted, and remains at $(x, p)$
otherwise. \progg{hmc-harmonic} translates the example program of \REF{Neal2011}
into our context.

\begin{algorithm}
    \newcommand{\algo}{hmc-harmonic}
    \captionsetup{margin=0pt,justification=raggedright}
    \begin{center}
        $\begin{array}{ll}
            & \PROCEDURE{\algo}\\
            & \INPUT{\SET{x_0 \TO  x_{N-1}},  t}\
            \COMMENT{configuration $x$, time}\\
            & \IS{\SET{\xtilde_0 \TO  \xtilde_{N-1}}} {\SET{x_0 \TO x_{N-1}}} \\
            &  \IS{\SET{p_0 \TO p_{N-1}}}
              {\SET{\sub{gauss}(0,1) \TO \sub{gauss}(0,1)}} \\
            & \IS{E}{U(x) + \half \sum_i p_i^2} \\
            &  \IS{\SET{p_0 \TO p_{N-1}}}
              {\SET{p_0 - \half \epsilon \nabla_0 U(\xtilde) \TO p_{N-1} -
\half\epsilon
              \nabla_{N-1} U(\xtilde)}}\ \COMMENT{start leapfrog}\\
            & \FOR{\iota \in \SET{0, I-1}}\\
            & \BRACE{
              \IS{\SET{\xtilde_0 \TO \xtilde_{N-1}}}
              {\SET{\xtilde_0 + \epsilon p_0 \TO \xtilde_{N-1} + \epsilon
p_{N-1}}} \\
            \IF{\iota \neq I-1}
              \IS{\SET{p_0 \TO p_{N-1}}}
              {\SET{p_0 - \epsilon \nabla_0 U(x) \TO p_{N-1} - \epsilon
              \nabla_{N-1} U(x)}} \\
                 } \\
            &  \IS{\SET{p_0 \TO p_{N-1}}}
              {\SET{p_0 - \half \epsilon \nabla_0 U(\xtilde) \TO p_{N-1} -
\half\epsilon
              \nabla_{N-1} U(\xtilde)}} \ \COMMENT{end leapfrog} \\
            & \IS{t}{t + N (2 I +1)}\ \COMMENT{set $\IS{p}{-p}$ to test
                           time-reversibility }\\
            & \IS{\Etilde}{U(\xtilde) + \half \sum_i p_i^2}\\
            & \IS{\Upsilon}{\ranb{0,1}}\\
          * & \IF{\Upsilon < \expc{- \glb \Etilde - E \grb }}
             \IS{\SET{x_0 \TO  x_{N-1}}} {\SET{\xtilde_0 \TO \xtilde_{N-1}}} \\
            & \OUTPUT{\SET{x_0 \TO  x_{N-1}}, t }\
\COMMENT{configuration $x'$ (either unchanged or $\xtilde$),  time}\\
            & \ENDPROCEDURE\
        \end{array}$
    \end{center}
\caption{\sub{\algo}. HMC for the harmonic
chain using the leapfrog algorithm (see \eq{equ:GradientHarmonic} for
the  gradient $\nabla_k U$. The algorithm depends on the stepsize
$\epsilon$ and the iteration number $I$, but not explicitly on $b$.}
\label{alg:\algo}
\end{algorithm}
\subsection{HMC acceptance rate, stability}
\label{subsec:HMCStability}

\begin{figure}[htb]
    \centering
    \includegraphics[width=1.0
\columnwidth]{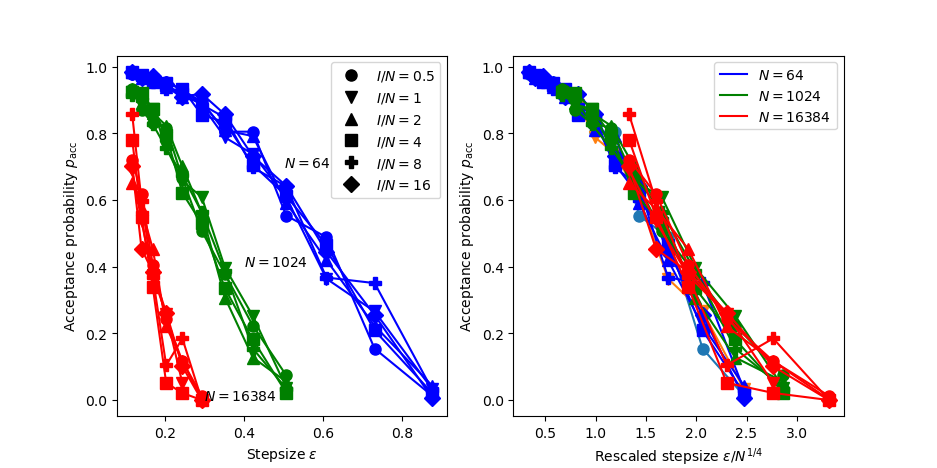}
    \caption{Stability of HMC for the harmonic chain.
    \subcap{Left} Acceptance probability $\pacc $
    of the Metropolis decision (line $*$ in \progn{hmc-harmonic}) \vs\
stepsize $\epsilon$ for different $N$
and ratios $I / N$, illustrating that HMC allows for large moves $x \to x'$.
\subcap{Right} Same data plotted \vs\ $N^{1/4} \epsilon$, suggesting $\sim
1/N^{1/4}$ scaling of $\pacc$.
}
\label{fig:HamiltonianAcceptanceProbability}
\end{figure}

Much research has gone into the study of
time-stepped dynamical systems~\cite{Leimkuhler2005} and the discretization
errors introduced by finite values of $\epsilon$. The key concept is that, with
a finite $\epsilon$, the discretized time evolution conserves the energy not of
the hamiltonian with energy $U$, but of a nearby \quot{shadow}
hamiltonian characterized by a potential energy $U^*$. As $U \sim U^*$, and
$U^* + K$ is conserved, the energy $U$ (as a function of the index $\iota$ in
\progn{hmc-harmonic}) remains close to $U^*$, allowing  HMC to
propose very large moves. The shadow hamiltonian usually describes the dynamics
up to large timescales, but it hypothetical. Asymptotically, at very
large times $\iota \epsilon$, the discretized dynamics  gives rise to a
drift~\cite{Engle2005}. However, for quadratic hamiltonians as the harmonic
chain, the shadow hamiltonian actually
exists~\cite{Neal2011,Pastor1988,Skeel2001}. As a consequence, the acceptance
probability of the Metropolis filter in \prog{hmc-harmonic} is largely
independent of $I$ for given $\epsilon$, allowing for non-local moves $x \to x'$
whereas it vanishes exponentially with $\delta$ for \prog{metropolis-harmonic},
keeping the moves $x \to x'$ local. This essential advantage of HMC is
illustrated in \fig{fig:HamiltonianAcceptanceProbability}. It survives up to
very long times for general hamiltonians.

For fixed $N$,  the HMC algorithm reaches large moves, but the acceptance
probability degrades with increasing stepsize, and it is essentially zero for
large $\epsilon$. For large $N$, this problem becomes more pronounced. For a
scaling $\epsilon \sim 1/N^{1/4}$, a universal acceptance probability is reached
(see \fig{fig:HamiltonianAcceptanceProbability}). This behavior is well
understood theoretically~\cite{Neal2011}.

\subsection{Relation of HMC to molecular dynamics}
\label{subsec:HMCBasicRequirements}
HMC, as implemented in \prog{hmc-harmonic}, contains three essential ingredients
that are associated with distinct computations. The first ingredient is the
repeated evaluation of the gradient which iteratively time-steps the leapfrog
algorithm from $x$ (and its associated momentum $p$)  to $x'$ (and $p'$). In the
harmonic chain, this step is insensitive to the parameter $b$, and its
computational cost is \bigOb{N} operations per time step (that is, per iteration
$\iota$ in \progn{hmc-harmonic}). With long-range
interactions, the evaluation of the gradient typically relies on fast Ewald
methods~\cite{hockney1988ComputerSimulationUsingParticles}
or multipole expansions~\cite{Greengard1987} and cannot be performed
efficiently to
machine precision~\cite{Hoellmer2024}. The second ingredient is the Metropolis
filter that
enforces the correct stationary distribution. It renders HMC reversible and
rigorous. It evaluates the total-energy change between $(x,p)$
and $(\xtilde, \ptilde)$, and is again independent of $b$. In the general
case,
ECMC avoids the evaluation of  $U$ or its derivatives yet still samples
$\pi = \expb{-\beta U}$ (see \REF{Tartero2024b} for an introduction). The
third ingredient is the refreshment (or thermalization) of the momenta---on the
same time frame as the Metropolis step, which assures the irreducibility of the
Markov chain. ECMC, as we will discuss later, also features auxiliary variables
that are updated as the simulation proceeds, but never refreshed.

With its rejection step, HMC samples the stationary distribution exactly.
It is thus a rigorous variant of molecular dynamics, the
hamiltonian time evolution coupled to a thermostat at all time steps $\epsilon$,
but without a
Metropolis decision.
Such a thermostat hides instabilities
(as shown in \fig{fig:HamiltonianAcceptanceProbability}), accumulated errors
and energy drifts. Thermostats thus may
produce artifacts, but they often appear as a
\quot{necessary evils}~\cite{Wong2016}. ECMC evaluates neither
gradients not differences of the total energy and requires no refreshments
and no thermostats. Its dynamics, however, claims no similarity with the
physical dynamics.

\section{Event-chain Monte Carlo for the harmonic chain}
\label{sec:ECMC}

For the harmonic chain, ECMC~\cite{Bernard2009,Michel2014JCP} corresponds to the
factorized Metropolis algorithm of \progn{metropolis-factorized-harmonic}
(or its variants) with
three changes. First, only forward moves are proposed.
This is feasible because of the periodic boundary conditions, but can be
easily generalized. The
\emph{a priori} probability $\ACAL(x,x')$ of \eq{equ:APrioriSymmetric} is thus
non-symmetric. Second, moves are infinitesimal, resulting in a continuous-time
Markov process. The rejection probability is so small
(it is infinitesimal), that most moves are accepted. Any rejection
can be traced to a specific factor,
and in \progn{metropolis-factorized-harmonic}
either to the forward factor (the one involving $U_k^+$)
or to the backward factor (the one
involving $U_k^-$), but not to both. Third, rejections are replaced by
\quot{liftings}. This means that, after an accepted (infinitesimal) move of
particle $k$, the same particle $k$ moves again. Otherwise, if $k$'s  move  is
rejected by $k^+$, $k^+$ will move next. If $k$'s move is rejected by $k^-$, it
is $k^-$ that will move next. The algorithm thus differs from
the reversible local algorithms
(\progn{metropolis-harmonic} and \progn{metropolis-factorized-harmonic}) in that
the particle to be moved at time $t+ \Delta t$ is computed
from the outcome of the simulation at time $t$. The particle index $k$ is
part of an enlarged \quot{lifted} sample space. Lifted
Markov chains~\cite{Diaconis2000,Chen1999} build the conceptual underpinning of
ECMC, a topic developed elsewhere~\cite{Krauth2021eventchain}.

In \SECT{subsec:TwoFactorECMC}, we implement ECMC for the harmonic
chain, as was already done in \REF{Lei2019}, starting from the
factorized filter of
\eq{equ:FactorizedFilter}
for infinitesimal proposed moves $x_k' = x_k + \diff x$. We then
focus on the role of the pressure of this algorithm. In
\SECT{subsec:FourFactorECMC}, we discuss other factorizations, in particular
writing the energy as $U = \sum(x_k - x_{k-1} )^2 - b \sum(x_k - x_{k-1}) +
\const$ (that is, following \eq{equ:UofxFactorFields}
rather than \eq{equ:UofxB}), which then
allows for generalizations to arbitrary interactions.
Finally, in \SECT{subsec:GeneralECMC}, we discuss general properties of
ECMC for the harmonic chain and its generalizations.

\subsection{Implementation of ECMC, role of pressure}
\label{subsec:TwoFactorECMC}

For the harmonic chain, ECMC may be implemented with the decomposition $U =
\half \sum(x_k - x_{k-1} -b)^2$ of \eq{equ:UofxB}.
An infinitesimal forward move of particle $k$ may change two terms in the above
sum, giving rise to two independent factor filters:
\begin{equation}
\begin{aligned}
\pFactp_k(x_k,  x_k + \diff x) = 1 -\half \maxc{0,\fracb{\partial (x^+ -
x_k -  b)^2}{\partial
x_k} }\diff x_k \\
\pFactm_k(x_k, x_k + \diff x)  = 1 - \half \maxc{0,\fracb{\partial U(x_k -
x^-  b)^2}{\partial x}
}\diff x_k \\
\end{aligned}
\end{equation}
For the eponymous event-driven implementation of ECMC, the time-and-space
interval $\diff t$ is not discretized~\cite{Peters_2012}. This is possible
because, after an accepted move of particle $k$, the same particle moves again.
For both factors, \quot{downhill} moves are always accepted and \quot{uphill}
moves are accepted until the total energy change equals the logarithm of a
uniform random number (see \REF[]{Tartero2024} for an introduction). For each
factor, we thus accept the downhill motion (to positions $z^\pm$ in
\progn{ecmc-harmonic}), then compute \quot{candidate}
events~\cite{Faulkner2018,Hoellmer2020}, for each of the factors, and finally
terminate the uphill move at the earliest of the candidate events, as it breaks
the consensus at the heart of the factorized Metropolis filter. After the
event with the active particle $k$, the particle $ k^+$ becomes active if the
event involves $\pFactp$, and the particle $ k^-$ if it involves
$\pFactm$.

\begin{algorithm}
    \newcommand{\algo}{ecmc-harmonic}
    \captionsetup{margin=0pt,justification=raggedright}
    \begin{center}
        $\begin{array}{ll}
            & \PROCEDURE{\algo}\\
            & \INPUT{\SET{x_0 \TO \vv{x_k} \TO x_{N-1}}, t}\ \COMMENT{lifted
configuration, time}\\
            & \IS{k^+}{\sub{mod}(k+1,N)},\
             \IS{k^-}{\sub{mod}(k-1,N)} \\
            & \IS{x^+}{x_{k^+}},\ \IF{k = N-1}{\IS{x^+}{x^+ + L}} \\
            & \IS{x^-}{x_{k^-}},\ \IF{k = 0}{\IS{x^-}{x^- - L}} \\
            & \IS{z^\pm}{x^\pm \mp b}\ \COMMENT{ends of downhill
motion}\\
            & \IS{\Upsilon^\pm}{-2 \logc{\ranb{0,1}}}\
            \COMMENT{two independent random numbers limiting uphill motion} \\
            & \IF{x_k < z^+ }\ \IS{\Delta^+}{z^+ -x_{k} + \sqrt{\Upsilon^+}}\
             \ELSE\ \IS{\Delta^+}{\sqrt{\Upsilon^+ + (x_k - z^+) ^2} + z^+ -
x_k }\\
            & \IF{x_k < z^- }\ \IS{\Delta^-}{z^- -x_{k} + \sqrt{\Upsilon^-}}\
             \ELSE\ \IS{\Delta^-}{\sqrt{\Upsilon^- + (x_k - z^-) ^2} + z^- -
x_k }\\
        * & \IF{\Delta^+ < \Delta^-}\ \IS{x_k}{x_k + \Delta^+},\
\IS{k}{k^+},\
                  \IS{t}{t + \Delta^+} \\
        \#   & \ELSE\ \IS{x_k}{x_k + \Delta^-},\
            \IS{k}{k^-},\
            \IS{t}{t + \Delta^-} \\
            & \OUTPUT{\SET{x_0 \TO \vv{x_k} \TO x_{N-1}}, t }\
\COMMENT{lifted configuration, time}\\
            & \ENDPROCEDURE\
        \end{array}$
    \end{center}
\caption{\sub{\algo}. Harmonic-chain ECMC (factors from $U = \sum_jk (x_{k+1} -
x_k - b)^2$). If $\Delta^+< \Delta^-$ (line $*$), the activity moves from the
initial $x_k$ to $x^+$ in time $\Delta^+$. Otherwise (line $\#$), it moves from
the the initial $x_k$ to  $x^-$ in time $\Delta^-$. For sampling, the algorithm
must be interrupted at equally spaced intervals (see the Python implementation
in \app{app:Algorithms}).}
\label{alg:\algo}
\end{algorithm}

\prog{ecmc-harmonic} implements ECMC, moving forward from one event to the next.
In practice, the algorithm may be stopped at regular time intervals, say, after
a time $\tau$ (see the Python implementation of \progn{ecmc-harmonic}).
At any generic time $t$,
the \quot{pointer} $\vv{x}$ (the position
of the
active particle $k$) moves forward with $x_k$, but at an event, it jumps
to the position  $x^+$ or  $x^-$. These jumps can be in both directions.
In a time $\tau$,
the pointer moves on average by a distance $\vpmean \tau$.
The mean pointer speed $\vpmean$
is linearly related to the system pressure~\cite{Michel2014JCP}:
\begin{equation}
   \beta P = \frac{N}{L } \vpmean.
\label{equ:FabulousFormulaTest}
\end{equation}
It follows from \eq{equ:FabulousFormula} that
\begin{equation}
 \vpmean =
 \frac{1}{\rho} \glb b  + \frac{1}{L }- \frac{1}{\rho} \grb
 = \frac{1}{\rho} \glb b  - \bcrit \grb
\label{equ:RhoRhoCrit}
\end{equation}
with the global density $\rho = N/L$ and with $\bcrit$ as defined earlier (see
\tab{tab:Pressure}). We may
interpret this equation in two ways. First, for fixed global density $\rho$,
the pointer velocity is obviously a linear function of $b$, hitting zero at
$\bcrit$. Second, for
$b$ fixed at $1/\rho$ (neglecting the $1/L$ terms),  local coarse-grained
densities $\rho + \Delta \rho$ will lead to a (signed) local pointer
velocity~\cite{Erignoux2024}:
\begin{equation}
 \vpointer \sim  \frac{1}{\rho^2} \glb \frac{\Delta \rho}{\rho}  \grb.
\label{equ:RhoRhoCritBrune}
\end{equation}
In regions with an excess density,  the pointer thus moves forward,
and in regions with a density deficit, it  moves backwards, getting trapped
between any two such regions~\cite{Erignoux2024}.

\begin{table}[]
    \centering
    \begin{tabular}{ccc}
        $b$ &  $\vpmean$ (pointer velocity)&
\eq{equ:RhoRhoCrit} \\ \hline
$1.7$ &$ -0.40014 \pm  0.0002$ & $-0.4$ \\
$1.8$ &$ -0.19991 \pm  0.0003$ & $-0.2$ \\
$1.9$ &$ -0.00020 \pm  0.0002$ & { }$0.0$ \\
$2.0$ { }&$  0.19998 \pm  0.0002$ & { }$0.2$ \\
$2.1$ { }&$ 0.39964 \pm  0.0003 $ & { }$0.4$
    \end{tabular}
     \caption{ECMC pointer velocity for the harmonic chain
    ($N=5$, $L=10$, $\beta=1$). Simulation results from
\prog{ecmc-harmonic}
are compared to the values derived from the known pressure
(\eq{equ:FabulousFormula}).
}
\label{tab:Pressure}
\end{table}

\subsection{Facter-field ECMC interpretation}
\label{subsec:FourFactorECMC}

The factorization used in \prog{ecmc-harmonic}
is by no means unique. An alternative is suggested by the
decomposition $U = \half \sum (x_k - x_{k-1})^2 -h \sum (x_k -
x_{k-1}) + \const$ (see \eq{equ:UofxFactorFields}), already implemented within
the local reversible
factorized Metropolis algorithm (see \progn{metropolis-fourfactor-harmonic}).
The linear term in this decomposition, introduced in \REF{Lei2019} as a
\quot{factor field}, sums up to zero because of the periodic boundary
conditions, and can be added to any continuous potential. In the implementation
of  \prog{ecmc-fourfactor-harmonic}, a move of a single particle
$x_k \to x_k'$ changes four factors, and (naively) calls for the consensus of
four decisions, of which one is to systematically accept the move.
Remarkably, the mean pointer velocity of
\progn{ecmc-fourfactor-harmonic} is the same as that of
\progn{ecmc-harmonic}.

\begin{algorithm}
    \newcommand{\algo}{ecmc-fourfactor-harmonic}
    \captionsetup{margin=0pt,justification=raggedright}
    \begin{center}
        $\begin{array}{ll}
            & \PROCEDURE{\algo}\\
            & \INPUT{\SET{x_0 \TO \vv{x_k} \TO x_{N-1}}, t, b}\
\COMMENT{lifted
configuration, time, suppose $b>0$}\\
            & \IS{k^+}{\sub{mod}(k+1,N)},\
             \IS{k^-}{\sub{mod}(k-1,N)} \\
            & \IS{x^+}{x_{k^+}},\ \IF{k = N-1}{\IS{x^+}{x^+ + L}} \\
            & \IS{x^-}{x_{k^-}},\ \IF{k = 0}{\IS{x^-}{x^- - L}} \\
            & \IS{\Upsilon^\pm_h}{-2 \logc{\ranb{0,1}}};\
             \IS{\Upsilon^+_f}{- \logc{\ranb{0,1}}}\
            \COMMENT{three independent random numbers} \\
            & \IF{x_k < x^+ }\ \IS{\Delta^+}{x^+ -x_{k} + \sqrt{\Upsilon^+_h}}
\\
            & \ELSE\ \IS{\Delta^+}{x^+ -x_{k} + \sqrt{\Upsilon^+_h + (x_k
-x^+) ^ 2}}\\
            &\IS{\Delta^+_f}{\Upsilon^+_f / b} \\
            & \IF{x_k < x^- }\ \IS{\Delta^-_h}{x^- -x_{k} +
\sqrt{\Upsilon^-_h}}\\
            & \ELSE\ \IS{\Delta^-_h}{x^- -x_{k} + \sqrt{\Upsilon^-_h + (x_k -
x^-)^2}}\\
            & \IS{\Delta^-} {\minb{\Delta^-_h, \Delta^-_f}} \\
            & \IF{\Delta^+ < \Delta^-}\ \IS{x_k}{x_k + \Delta^+},\ \IS{k}{k^+},\
                  \IS{t}{t + \Delta^+} \\
            & \ELSE\ \IS{x_k}{x_k + \Delta^-},\
            \IS{k}{k^-},\
            \IS{t}{t + \Delta^-} \\
            & \OUTPUT{\SET{x_0 \TO \vv{x_k} \TO x_{N-1}}, t }\
\COMMENT{lifted configuration, time}\\
            & \ENDPROCEDURE\
        \end{array}$
    \end{center}
    \caption{\sub{\algo}.
    Harmonic-chain ECMC (factors from $U = \half \sum (x_k - x_{k-1})^2 -h \sum
(x_k - x_{k-1}) + \const$). This algorithm tracks more factors
than \progn{ecmc-harmonic}, but it samples the same Boltzmann distribution.}
\label{alg:\algo}
\end{algorithm}

\subsection{Generalization of ECMC}
\label{subsec:GeneralECMC}

ECMC implements a non-reversible Markov chain in the lifted sample space
composed of the particle configuration and the identity of the active particle.
For all values of the \equidist it  is irreducible and aperiodic. With its
constant velocity (in this simplest setting), it does not conserve the energy
$U$, and the event-driven formulation sidesteps all the issues related to the
discretization of time-driven dynamics. Finally, there are no thermostats.
The ECMC algorithms studied here have been generalized in different directions.
Newtonian ECMC, for example, ascribes velocities to all particles, but actually
moves only a subset of them~\cite{Klement2019}. This then
parallels the dynamics of the HMC, but it remains event-driven
and non-reversible at all times and exactly samples the distribution $\pi$.
Variants of ECMC may~\cite{Bernard2009} or may not~\cite{Michel2020} require
refreshments~\cite{Hoellmer2022}. ECMC is
embedded in the framework of piecewise deterministic Markov
processes~\cite{Davis1984,BierkensPDMC2017,Bierkens2019}.

\section{Comparison of MCMC algorithms for the harmonic chain}
\label{sec:ComparisonsBenchmarks}
In \SECT{subsec:DefinitionsThermodynamics}, we analytically obtained the
thermodynamics of the harmonic string, in other words its steady-state behavior
towards which all discussed algorithms converge rigorously. The complete
independence of results on the step size  $\epsilon $ and the number of
iterations $I$ (for HMC) and on the equilibrium distance $b$ (for ECMC) is
checked explicitly in \SECT{subsec:BenchmarkDefinitions},
testing  our implementations. HMC depends on two
external parameters, namely the number $I$ of iterations and the stepsize
$\epsilon$, while ECMC depends on  the \equidist. Following \REF{Neal2011} for
HMC and \REF{Lei2019} for ECMC, we discuss in
\SECT{subsec:RefreshmentsFineTuning} how to choose optimal parameters. Finally,
in \SECT{subsec:BenchmarkRelated}, we discuss the scaling of autocorrelation
times with system size.

\subsection{Mean energy}
\label{subsec:BenchmarkDefinitions}

As a first check of our algorithms, we determine the $b$-independent
part of the mean energy $\mean{U}$. Five-digit precisions are obtained easily,
and the agreement with the analytical result of \eq{equ:MeanEnergy} is complete.
All the MCMC computations start from an equilibrium configuration obtained by
the Lévy construction, so that no error is introduced by the approach of the
steady state (see \tab{tab:Energy}). The absence of discretization errors is
particularly remarkable. HMC, as discussed, propagates a shadow hamiltonian
rather than the one corresponding to the energy $U$. It is the
Metropolis rejection step that enforces detailed balance with respect to the
correct Boltzmann distribution. In ECMC, all rejections are, so to speak,
re-branded into lifting moves, in our case the transfer of the pointer from one
particle to another. Its event-driven setup, by definition, avoids all
time-discretization.  All Python programs are openly available
(see \app{app:Access}).

\begin{table}[]
    \centering
    \begin{tabular}{ccc}
        Method &  $\mean{U}$ ($b$-independent part) & see \\ \hline
exact& $ 19.5 = \half( L^2 / N + N-1)$ & \eq{equ:MeanEnergy} \\
Lévy construction & $ 19.50025 \pm 0.0004$ & \prog{levy-harmonic} \\
Metropolis & $19.5002 \pm 0.0004$  & \prog{metropolis-harmonic} \\
Factorized  ($b=1$) & { }$19.49995 \pm  0.0003$ &
\prog{metropolis-factorized-harmonic} \\
Factor-field ($b= 2.5$) & { }$19.5002 \pm  0.006$
& \prog{metropolis-fourfactor-harmonic} \\
HMC ($\epsilon = 0.1, I=20$) &  $19.50043 \pm 0.0007$
& \prog{hmc-harmonic} \\
HMC ($\epsilon = 0.4, I=5$) & $ 19.49969 \pm
0.0006$     & \prog{hmc-harmonic} \\
ECMC ($b=1$) &  $19.50075 \pm 0.0008$ & \prog{ecmc-harmonic} \\
ECMC ($b=2$) &  $19.50017 \pm 0.0006$ & \prog{ecmc-harmonic}
    \end{tabular}
     \caption{Estimates of the $b$-independent part of the mean energy
$\mean{U}$ for $N=8$, $L=16$, from sampling methods discussed in this
paper. All algorithms rigorously sample the Boltzmann distribution of
\eq{equ:Boltzmann}.}
\label{tab:Energy}
\end{table}

\subsection{Parameter dependence of HMC and ECMC}
\label{subsec:RefreshmentsFineTuning}

For a configuration $x = \SET{x_0 \TO x_{N-1}}$ at time $t$, we define the
structure factor as:
\begin{equation}
S(x,t) = \frac{1}{N} \gle \sum_{j=0}^{N-1}\expb{i q
x_j}\gre ^2,
\label{equ:StructureFactor}
\end{equation}
where $q = 2 \pi / L $ is the smallest non-zero wave number in a periodic system
of length $L$. The structure factor is sensitive to long-range density
fluctuations, which are expected to relax slowly in equilibrium. It is
invariant under uniform translations of all particles, and it does not require
to know particle indices or the connectivity matrix.
As in earlier works, we use the autocorrelation time of the structure factor to
benchmark MCMC algorithms, but a number of related observables give equivalent
results.

\begin{figure}[htb]
    \centering
    \includegraphics[width=7cm]{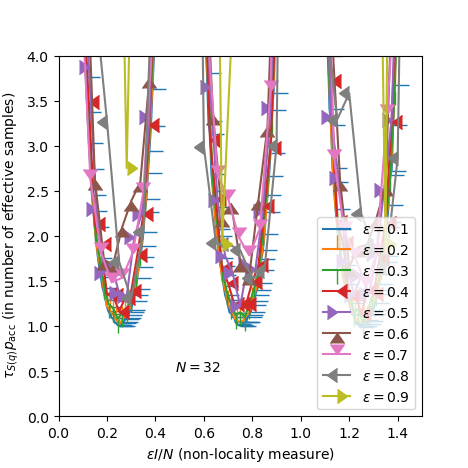}
    \includegraphics[width=7cm]{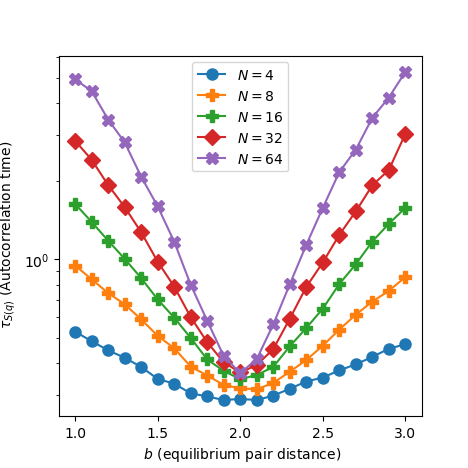}
    \caption{Parameter-dependence of the equilibrium dynamics
    for the harmonic chain at $L= 2N$:
    \subcap{Left} Autocorrelation time, measured in \emph{number of
\quot{chains}} multiplied by the acceptance rate of the Metropolis filter, for
HMC \vs\ effective chain length $\epsilon I$ rescaled by $N$, showing that a
\bigOb{1} accepted chains of length $\epsilon I / N \sim \const$ decorrelate the
model. \subcap{Right} Autocorrelation time for ECMC, rescaled by $N^{3/2}$,
featuring strong $b$ dependence (see \REF{Lei2019}).
}
\label{fig:AutocorrelationsFixedN}
\end{figure}

As discussed, HMC proposes non-local moves $x \to
x'$ from the Newtonian evolution of an associated dynamical system.
In the harmonic chain, these proposals are more efficient than those of the
local Metropolis algorithm as, for $\epsilon \to 0$, essentially a single long
move decorrelates the Markov chain. The length of this chain
requires fine-tuning to values $\epsilon I \simeq 0.25N, 0.75N, \dots$ in order
to avoid recurrence with $x \simeq x'$
(see \fig{fig:AutocorrelationsFixedN}). For small non-zero step sizes
$\epsilon$, the quality of the proposed moves degrades gracefully, and it
remains true that a single long, \emph{accepted}, move decorrelates the Markov
chain. The computing effort for one such move is $ \tau \sim N^{9/4}$, as we
have to choose $\epsilon \sim 1/N^{1/4}$ to keep the rejection rate under
control.

The performance of ECMC naturally depends on the \equidist (in more general
terms, on the factor field). The remarkable speedup of autorcorrelation times
at $b = \bcrit$, shown previously~\cite{Lei2019} in the harmonic chain and its
generalizations, is shown again in \fig{fig:AutocorrelationsFixedN}.

\subsection{Scaling of autocorrelation times}
\label{subsec:BenchmarkRelated}

\begin{figure}[htb]
    \centering
\includegraphics[width=9cm]{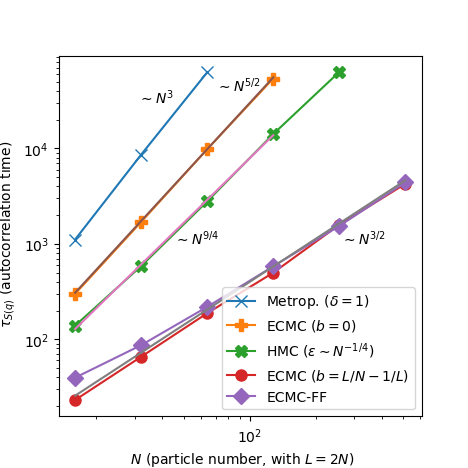}
    \caption{Autocorrelation times, in single moves or events, for the
structure factor in the harmonic chain with $L=2N$.
The diffusive relaxation in \bigOb{N^3} for the Metropolis
algorithm, contrasts with the \bigOb{N^{5/2}} behavior for unadjusted
ECMC (for $b \neq \bcrit$) and
that the \bigOb{N^{9/4}} scaling for HMC, which can be decomposed into
\bigOb{N^2} to move
every particle by $\bigOb{L}$, and \bigOb{N^{1/4}} for the choice of the
stepsize $\epsilon$, which controls the instability
of \fig{fig:HamiltonianAcceptanceProbability}. The two versions of
ECMC correspond to \progn{ecmc-harmonic} and \progn{ecmc-fourfactor-harmonic}.
For $b = \bcrit$,  they both scale as \bigOb{N^{3/2}}.
}
\label{fig:AutocorrelationsHmcECMC}
\end{figure}

We again consider the harmonic chain of $N$ particles with $L=2N$, initialized
in equilibrium through a  direct sample of \prog{levy-harmonic}.
The data indicate that a local reversible Markov chain, here
\prog{metropolis-harmonic}, relaxes in \bigOb{N^3} single moves (see
\fig{fig:AutocorrelationsHmcECMC}). This corresponds to \bigOb{N^2}
\quot{sweeps} of $N$ particles, in other words a dynamical scaling with exponent
$\mu = 2$. This scaling is characteristic of models in the
Edwards--Wilkinson class~\cite{Edwards1982surface}. Relaxation in \bigOb{N^3}
single steps is proven~\cite{Caputo2010} rigorously for the symmetric simple
exclusion model (SSEP), a discrete lattice reduction of the harmonic chain. The
ECMC algorithm, for an unadjusted equilibrium pair distance $b \neq \bcrit$,
features a relaxation in \bigOb{N^{5/2}} single moves (or
\bigOb{N^{3/2}} sweeps). This behavior was identified, in a related lattice
model~\cite{KapferKrauth2017}, as representing  the universality class of the
TASEP described by the
Kardar--Parisi--Zhang (KPZ) exponent $3/2$. For unadjusted pullback $\alpha$,
the lifted TASEP likewise has an \bigOb{N^{5/2}} inverse gap~\cite{Essler2024}.
HMC (with a stepsize $ \epsilon \sim 1/N^{1/4}$) and a chain length $\epsilon I
\simeq 0.25 N$ is very well compatible with the expected
\bigOb{N^{9/4}} scaling behavior. Finally, the
\bigOb{N^{3/2}} scaling  of the autocorrelation time at $b = \bcrit$ confirms
analogous prior computations in the same model~\cite{Lei2019}.
The \bigOb{N^{3/2}} decay of autocorrelations  was also found in the lifted
TASEP~\cite{Essler2024}. The behavior is generic, and not related to the
integrability of the Gaussian model and the lifted TASEP.  Nevertheless,
there is very strong evidence, from the Bethe ansatz, of an inverse gap scaling
as \bigOb{N^2} in that model.

\section{Discussion, conclusion}
\label{sec:Conclusion}

Reversible, local Markov chains form the backbone of the huge field of Monte
Carlo methods. The reversibility of these methods brings in the detailed-balance
condition, with which a move $x \to x'$ can be accepted or rejected using a
filter involving only $x$ and $x'$. As we have discussed, the locality of these
methods is the price to be paid for the move $x \to x'$ being generated blindly,
in our case by a random displacement of a particle. Large moves exit the region
of sample space with high weight (low energy), and get rejected. Local-move
reversible MCMC algorithms for one-dimensional particle models generally have
a diffusive relaxation time \bigOb{N^3}, and this can often be proven
rigorously.

To propose non-local moves $x \to x'$, HMC relies on \quot{[g]eometric
integrators [that] are time-stepping methods, designed so that they exactly
satisfy conservation laws, symmetries, or symplectic properties of a system of
differential equations} (quote from \REF{Leimkuhler2005}). This results in
custom-tailored, non-local moves, but which are costly to construct. In the
harmonic chain, the leapfrog integrator of Newton's equations comes with a time
complexity \bigOb{N^{9/4}}, of which \bigOb{N^{1/4}} control the numerical
instability of the geometric integrator, and \bigOb{N^2}  moves each of the
particles ballistically by a distance on the order of the system size. The
\bigOb{N^{9/4}} time complexity (corresponding to \bigOb{N^{5/4}} sweeps) is
only slightly larger than the $\mu = 1$ dynamical scaling exponent
associated with molecular dynamics. It is generic for a wide
range of one-dimensional particle models, geometric integrators and differential
equations.

To propose moves $x \to x'$, ECMC relies on a class of event-stepping methods
designed to exactly satisfy the very global-balance condition that encodes the
stationarity of the Boltzmann distribution. The method has no rejections
because, when a particle
displacement is impossible, a \quot{lifting} move takes place. Past research has
shown this paradigm to adapt itself to real-life models in statistical
mechanics~\cite{Bernard2011} and chemical physics~\cite{Hoellmer2024}, with
demonstrated speedups of several orders of magnitude~\cite{Bernard2011,Li2022}.
The harmonic chain  demonstrates in a rather complicated setting
that non-physical local
dynamics can decorrelate even faster than the hamiltonian dynamics of HMC. This
point was made for molecular dynamics rather than HMC, in \REF{Lei2019}.

What we have discussed here is in no way related to any special
feature of the
harmonic model. The \bigOb{N^{3/2}} scaling of autocorrelation functions at the
critical factor field was observed also for Lennard-Jones interactions, and for
hard spheres in the continuum. Furthermore, it is also a definite
feature of the lifted TASEP~\cite{Essler2024}, where the Bethe ansatz yields
direct access to the excitation spectrum of the transition matrix.
In both the harmonic chain and the lifted TASEP, the superior performance
requires fine-tuning to vanishing pressure.
However,
this is easy, as the zero-pressure point corresponds to driftless pointer
motion.
In the lifted TASEP, there is compelling evidence for the
existence, in any momentum sector,
of a possibly
isolated eigenvalue with an
\bigOb{N^2} inverse-gap scaling~\cite{Essler2024}.
The corresponding state seems not to couple to \quot{normal}
observables~\cite{Essler2024b}.
It is unknown whether such a state is present in the
harmonic chain. The \bigOb{N^{3/2}} scaling at $\bcrit$ has natural support
from a hydrodynamic viewpoint~\cite{Erignoux2024}.

To make further progress, it appears important to study a possible
synthesis between HMC and ECMC, for example whether modifications of the
former can reach the \bigOb{N^{3/2}} scaling in the
harmonic chain.
In one-dimensional particle systems, it appears possible to
fully understand the remarkable ECMC dynamics.
The recently discovered deep
connection between ECMC and the fully self-avoiding random
walk\cite{Maggs2024,Maggs2024b} merits full attention.
Factor-field ECMC has also been implemented in
higher-dimensional particle systems, where it relaxes density modes with great
efficiency~\cite{MaggsKrauth2022}. In more than one dimensions, the dynamics of
non-reversible Markov chains needs to be much better characterized,
and the recent focus on one-dimensional systems should be seen as a first step.

\section*{Acknowledgments}
It is a pleasure to thank Nawaf Bou-Rabee, Bob Carpenter, Fabian Essler, and A.
C. Maggs for inspiring discussions, and the Simons Center (Flatiron Institute)
for hospitality. This research was supported by a grant from the Simons
Foundation (Grant 839534, MET).

\appendix
\section{Mathematical details}
\label{app:Math}

We compute the partition function of the harmonic chain, retaining the
dependence on $\beta$ in order to compute the internal energy.
The partition function of the model defined by
\eqfromto{equ:UofxB}{equ:UofxIndependent} is
\begin{align}
Z(N, b, L) &=
\expb{\beta b L - \half \beta  b^2 N}
 Z(N, 0, L)  \\
 &=
\expb{\beta b L - \half \beta  b^2 N}
 \int_{0}^L \diff x_0 \dots \int_{-\infty}^{\infty}\diff x_{N-1}
 \expc{- \beta U(x, 0, L)},
\label{equ:xxxUofxIndependent}
\end{align}
with $x_N = x_0 +L $ understood.
In \eq{equ:xxxUofxIndependent}, the exponential contains terms
$\half (x_k - x_{k-1})^2$ with $x_N = x_0 + L$. We may
change the boundary conditions to eliminate $L$
through the transformation:
\begin{align}
 y_k &= x_k - k \frac{L}{N}\quad k \in \SET{0 \TO N} \\
 y_{k-1} &= x_{k-1} - (k-1) \frac{L}{N},
\end{align}
resulting in:
\begin{align}
Z(N, b, L) &= \expc{\beta b L - \half \beta b^2 N - \half \beta L^2 N } \times
\\
&L \int_{-\infty}^{\infty} \diff y_1 \dots
\int_{-\infty}^{\infty} \diff y_{N-1} \expc{- \frac{\beta}{2} (y_k -
y_{k-1})^2}.
\label{equ:SecondIntegralZ}
\end{align}
Here, the prefactor $L$ comes from the integration of  $y_0$, which we now fix
to $0$. The coefficient $\beta$ can be absorbed in a rescaling
of the $y_k = \ytilde_k/\sqrt{\beta}$, so that the integral in
\eq{equ:SecondIntegralZ} becomes
\begin{equation}
 \frac{1}{\beta^{(N-1)/2}} \int \diff \ytilde_1 \cdots \diff \ytilde_{N-1}
\expb{-\half \sum_{i,j}
 A_{ij} \ytilde_i \ytilde_j } = \frac{1}{\beta^{(N-1)/2}}
 \sqrt{ \frac{(2\pi)^{N-1} }{\det A }},
\end{equation}
with an $(N-1) \times (N-1)$ Toeplitz matrix
(setting $\ytilde_0 = \ytilde_N = 0$)
\begin{equation}
A_{N-1} =
\begin{pmatrix}
2       & -1 & 0 & \cdots & \cdots& 0  \\
-1 & 2  & -1  &  0 & \cdots & \cdots\\
0 & -1 & 2  & -1  & 0 &  \cdots  \\
\vdots  & \vdots  & \ddots & \vdots & \vdots  \\
0& \cdots& \cdots & 0  & -1 & 2
\end{pmatrix};
\quad \det A_{N-1} = N
\end{equation}
leading to \eq{equ:PartitionFunction}. For the logarithm of the partition
function, we have
\begin{equation}
 \loga{Z(N, b, L)} = \loga{L} + \beta b L - \half \beta N b^2 - \half \beta L^2
/N
 + \const - \frac{N-1}{2} \loga{\beta}
\end{equation}
which implies \eqtwo{equ:FabulousFormula}{equ:MeanEnergy} for the pressure and
the mean energy.

\section{Sampling algorithms, implementations, website}
\label{app:Algorithms}

For the sake of completeness, we present two more sampling algorithms, one
being a local
Gibbs sampler, and the other the local reversible four-factor Metropolis
algorithm, which serves as a starting point for the factor-field variant of ECMC
(\prog{ecmc-fourfactor-harmonic}).

\subsection{Additional pseudocode algorithms}

In a heat-bath algorithm (Gibbs sampler), a subsystem  of the
configuration $x$ is equilibrated with respect to its environment.
The harmonic chain allows this subsystem to be the position of a single
particle $k$ or any number of neigboring particles. This subsystem can be
equilibrated through a variant of the Lévy construction, so that the Markov
chain (implemented in \prog{gibbs-harmonic}) need not be local. This contrasts
with the situation for the local Metropolis algorithm, where the step-size
$\delta$ and the number of displaced particles must be small in order to keep
a sizeable acceptance rate. In general models, the heat-bath and Metropolis
algorithms behave similarly.

\begin{algorithm}
    \newcommand{\algo}{gibbs-harmonic}
    \captionsetup{margin=0pt,justification=raggedright}
    \begin{center}
        $\begin{array}{ll}
           & \PROCEDURE{\algo}\\
           & \INPUT{\SET{x_0 \TO x_{N-1}}, t}\
            \COMMENT{configuration, time}\\
           & \IS{k}{\sub{choice}(\SET{0 \TO N-1})}\\
           & \IS{k^+}{\sub{mod}(k+1,N)},\
             \IS{k^-}{\sub{mod}(k-1,N)} \\
           & \IS{x^+}{x_{k^+}},\ \IF{k = N-1}{\IS{x^+}{x^+ + L}} \\
           & \IS{x^-}{x_{k^-}},\ \IF{k = 0}{\IS{x^-}{x^- - L}} \\
           & \IS{\overline{x}}{\half( x^+ + x^-)},\ \IS{\sigma}{1/\sqrt{2}}\\
           & \IS{x_k}{\sub{gauss}(\overline{x}, \sigma)}\ \COMMENT{compare to
\prog{levy-harmonic}}\\
           & \OUTPUT{\SET{x_0 \TO x_{N-1}}, t+1}\
\COMMENT{configuration, time}\\
           & \ENDPROCEDURE\
        \end{array}$
    \end{center}
    \caption{\sub{\algo}. Heatbath Monte Carlo algorithm for the harmonic chain.
The algorithm is independent of $b$.}
\label{alg:\algo}
\end{algorithm}

We provide the factorized Metropolis algorithm for the decomposition of the
potential into the factors indicated by
\eqtwo{equ:FourTermsInitial}{equ:FourTermsFinal}. The implementation is naive,
and the decision can be reduced from the consensus of four factors to that of
three factors (see \prog{metropolis-fourfactor-harmonic}). The algorithm is
closely related to the alternative \quot{factor-field} ECMC method
\prog{ecmc-fourfactor-harmonic}.

\begin{algorithm}
    \newcommand{\algo}{metropolis-fourfactor-harmonic}
    \captionsetup{margin=0pt,justification=raggedright}
    \begin{center}
        $\begin{array}{ll}
           & \PROCEDURE{\algo}\\
           & \INPUT{\SET{x_0 \TO x_{N-1}}, t}\
            \COMMENT{configuration, time}\\
           & \dots  \COMMENT{$k,\Delta_x, k^+,k^-, x^+ , x^-, x_k' $ as in
\progn{metropolis-harmonic}}\\
           & \IS{U_k^+}{\half  (x^+ - x_k)^2};\
            \IS{U_k^-}{\half  (x_k - x^-)^2}\\
           & \IS{U_k^{\prime +} }{\half  (x^+ - x_k')^2};\
            \IS{U_k^{\prime -}}{\half  (x_k' - x^-)^2}\\
           & \IS{U_f^+}{-b (x^+ - x_k)};\
            \IS{U_f^-}{ -b(x_k - x^-)}\\
           & \IS{U_f^{\prime +}}{-b (x^+ - x_k')};\
            \IS{U_f^{\prime -} }{ -b(x_k' - x^-)}\\
           & \IS{\Upsilon^+_k}{\ranb{0,1}};\  \IS{\Upsilon^-_k}{\ranb{0,1}}\\
           & \IS{\Upsilon^+_f}{\ranb{0,1}};\  \IS{\Upsilon^-_f}{\ranb{0,1}}\\
        *  & \IF{\Upsilon^+_k < \expc{-(U^{\prime +}_k- U^+_k)}\ \AND\
           \Upsilon^-_k < \expc{-(U^{\prime -}_k - U^{-}_k)} \  \\
           & \quad \AND\ \Upsilon^+_f < \expc{-(U^{\prime +}_f- U^+_f)}\ \AND\
           \Upsilon^-_f < \expc{-(U^{\prime -}_f - U^{-}_f)}}
           \IS{x_k}{x'_k} \\
           & \OUTPUT{\SET{x_0 \TO x_{N-1}}, t+1}\
\COMMENT{configuration, time}\\
           & \ENDPROCEDURE\
        \end{array}$
    \end{center}
    \caption{\sub{\algo}. Alternative factorized Metropolis algorithm for the
harmonic
chain. The two-line condition * checks for the consensus of the four factors
corresponding to \eqtwo{equ:FourTermsInitial}{equ:FourTermsFinal}. }
\label{alg:\algo}
\end{algorithm}

\subsection{Access to computer programs}
\label{app:Access}
This paper is accompanied by the \texttt{BeyondSamp} software
package,
which is published as an open-source project under the GNU GPLv3 license.
\texttt{BeyondSamp} is available on GitHub as a part of the JeLLyFysh
organization. The package contains Python scripts for the
algorithms
discussed in this paper and used to produce all the
numerical results. The url of the repository is
\url{https://github.com/jellyfysh/BeyondSamp.git}.


\begin{thebibliography}{10}

\bibitem{Metropolis1953}
N.~{Metropolis}, A.~W. {Rosenbluth}, M.~N. {Rosenbluth}, A.~H. {Teller}, and
  E.~{Teller}, ``{Equation of State Calculations by Fast Computing Machines},''
  {\em J. Chem. Phys.}, vol.~21, pp.~1087--1092, 1953.

\bibitem{Glauber1963}
R.~J. {Glauber}, ``{Time-Dependent Statistics of the Ising Model},'' {\em J.
  Math. Phys.}, vol.~4, pp.~294--307, 1963.

\bibitem{Creutz1980HeatBath}
M.~Creutz, ``{Monte Carlo study of quantized SU(2) gauge theory},'' {\em Phys.
  Rev. D}, vol.~21, pp.~2308--2315, 1980.

\bibitem{Geman1984}
S.~Geman and D.~Geman, ``{Stochastic Relaxation, Gibbs Distributions, and the
  Bayesian Restoration of Images},'' {\em IEEE Trans. Pattern Anal. Mach.
  Intell.}, vol.~{PAMI}-6, no.~6, pp.~721--741, 1984.

\bibitem{DuKePeRo1987}
S.~Duane, A.~D. Kennedy, B.~J. Pendleton, and D.~Roweth, ``{Hybrid
  Monte-Carlo},'' {\em Phys Lett B}, vol.~195, pp.~216--222, 1987.

\bibitem{Bernard2009}
E.~P. Bernard, W.~Krauth, and D.~B. Wilson, ``{Event-chain Monte Carlo
  algorithms for hard-sphere systems},'' {\em Phys. Rev. E}, vol.~80,
  p.~056704, 2009.

\bibitem{Michel2014JCP}
M.~{Michel}, S.~C. {Kapfer}, and W.~{Krauth}, ``{Generalized event-chain Monte
  Carlo: Constructing rejection-free global-balance algorithms from
  infinitesimal steps},'' {\em J. Chem. Phys.}, vol.~140, no.~5, p.~054116,
  2014.

\bibitem{Lei2019}
Z.~Lei, W.~Krauth, and A.~C. Maggs, ``{Event-chain Monte Carlo with factor
  fields},'' {\em Physical Review E}, vol.~99, no.~4, 2019.

\bibitem{Essler2024}
F.~H. Essler and W.~Krauth, ``{Lifted TASEP: A Solvable Paradigm for Speeding
  up Many-Particle Markov Chains},'' {\em Physical Review X}, vol.~14, no.~4,
  2024.

\bibitem{Neal2011}
R.~M. Neal, ``{MCMC using Hamiltonian dynamics},'' in {\em {Handbook of Markov
  Chain Monte Carlo}} (S.~Brooks, A.~Gelman, G.~Jones, and X.-L. Meng, eds.),
  pp.~113--162, Chapman and Hall/CRC, 2011.

\bibitem{SMAC}
W.~Krauth, {\em {Statistical Mechanics: Algorithms and Computations}}.
\newblock Oxford University Press, 2006.

\bibitem{RandallWinklerCircle2005}
D.~Randall and P.~Winkler, ``{Mixing Points on a Circle},'' in {\em
  Approximation, Randomization and Combinatorial Optimization. Algorithms and
  Techniques: 8th International Workshop on Approximation Algorithms for
  Combinatorial Optimization Problems, APPROX 2005 and 9th International
  Workshop on Randomization and Computation, RANDOM 2005, Berkeley, CA, USA,
  August 22-24, 2005. Proceedings} (C.~Chekuri, K.~Jansen, J.~D.~P. Rolim, and
  L.~Trevisan, eds.), pp.~426--435, Berlin, Heidelberg: Springer Berlin
  Heidelberg, 2005.

\bibitem{Levy1940}
P.~{L{\'e}vy}, ``{Sur certains processus stochastiques homog\`enes.},'' {\em
  {Compos. Math.}}, vol.~7, pp.~283--339, 1939.

\bibitem{Pollock1987}
E.~L. Pollock and D.~M. Ceperley, ``{Path-integral computation of superfluid
  densities},'' {\em Phys. Rev. B}, vol.~36, no.~16, pp.~8343--8352, 1987.

\bibitem{Krauth1996}
W.~Krauth, ``{Quantum Monte Carlo Calculations for a Large Number of Bosons in
  a Harmonic Trap},'' {\em Physical Review Letters}, vol.~77, no.~18,
  pp.~3695--3699, 1996.

\bibitem{Glasserman2003}
P.~Glasserman, {\em {Monte Carlo Methods in Financial Engineering}}.
\newblock Stochastic Modelling and Applied Probability, New York, NY: Springer,
  2003~ed., 2003.

\bibitem{Levin2008}
D.~A. Levin, Y.~Peres, and E.~L. Wilmer, {\em {Markov Chains and Mixing
  Times}}.
\newblock American Mathematical Society, 2008.

\bibitem{Tartero2024}
G.~Tartero and W.~Krauth, ``{Concepts in Monte Carlo sampling},'' {\em American
  Journal of Physics}, vol.~92, no.~1, p.~65–77, 2024.

\bibitem{Essler2024b}
F.~H. Essler, J.~Gipouloux, and W.~Krauth, 2024.
\newblock {Manuscript in preparation}.

\bibitem{Sinai1970}
Y.~G. {Sinai}, ``{Dynamical systems with elastic reflections},'' {\em Russian
  Mathematical Surveys}, vol.~25, pp.~137--189, 1970.

\bibitem{Simanyi2003}
N.~{Sim{\'a}nyi}, ``{Proof of the Boltzmann-Sinai ergodic hypothesis for
  typical hard disk systems},'' {\em Inventiones Mathematicae}, vol.~154,
  pp.~123--178, 2003.

\bibitem{Alder1957}
B.~J. {Alder} and T.~E. {Wainwright}, ``{Phase Transition for a Hard Sphere
  System},'' {\em J. Chem. Phys.}, vol.~27, pp.~1208--1209, 1957.

\bibitem{Alder1962}
B.~J. {Alder} and T.~E. {Wainwright}, ``{Phase Transition in Elastic Disks},''
  {\em Physical Review}, vol.~127, pp.~359--361, 1962.

\bibitem{Leimkuhler2005}
B.~Leimkuhler and S.~Reich, {\em {Simulating Hamiltonian Dynamics}}.
\newblock Cambridge: Cambridge University Press, 2005.

\bibitem{Engle2005}
R.~D. Engle, R.~D. Skeel, and M.~Drees, ``{Monitoring energy drift with shadow
  Hamiltonians},'' {\em Journal of Computational Physics}, vol.~206, no.~2,
  p.~432–452, 2005.

\bibitem{Pastor1988}
R.~W. Pastor, B.~R. Brooks, and A.~Szabo, ``{An analysis of the accuracy of
  Langevin and molecular dynamics algorithms},'' {\em Molecular Physics},
  vol.~65, no.~6, p.~1409–1419, 1988.

\bibitem{Skeel2001}
R.~D. Skeel and D.~J. Hardy, ``{Practical Construction of Modified
  Hamiltonians},'' {\em SIAM Journal on Scientific Computing}, vol.~23, no.~4,
  p.~1172–1188, 2001.

\bibitem{hockney1988ComputerSimulationUsingParticles}
R.~W. Hockney and J.~W. Eastwood, {\em {Computer Simulation Using Particles}}.
\newblock Bristol, PA, USA: Taylor \& Francis, Inc., 1988.

\bibitem{Greengard1987}
L.~Greengard and V.~Rokhlin, ``{A fast algorithm for particle simulations},''
  {\em J. Comput. Phys.}, vol.~73, no.~2, pp.~325--348, 1987.

\bibitem{Hoellmer2024}
P.~H\"{o}llmer, A.~C. Maggs, and W.~Krauth, ``{Fast, approximation-free
  molecular simulation of the SPC/Fw water model using non-reversible Markov
  chains},'' {\em Scientific Reports}, vol.~14, no.~1, p.~16449, 2024.

\bibitem{Tartero2024b}
G.~Tartero and W.~Krauth, ``{Fast sampling for particle systems with long-range
  potentials}.''
\newblock {Manuscript in preparation}.

\bibitem{Wong2016}
J.~Wong-ekkabut and M.~Karttunen, ``The good, the bad and the user in soft
  matter simulations,'' {\em Biochim. Biophys. Acta - Biomembr.}, vol.~1858,
  no.~10, pp.~2529--2538, 2016.

\bibitem{Diaconis2000}
P.~Diaconis, S.~Holmes, and R.~M. Neal, ``{Analysis of a nonreversible Markov
  chain sampler},'' {\em Ann. Appl. Probab.}, vol.~10, pp.~726--752, 2000.

\bibitem{Chen1999}
F.~Chen, L.~Lovász, and I.~Pak, ``{Lifting Markov Chains to Speed up
  Mixing},'' {\em Proceedings of the 17th Annual ACM Symposium on Theory of
  Computing}, p.~275, 1999.

\bibitem{Krauth2021eventchain}
W.~Krauth, ``{Event-Chain Monte Carlo: Foundations, Applications, and
  Prospects},'' {\em Front. Phys.}, vol.~9, p.~229, 2021.

\bibitem{Peters_2012}
E.~A. J.~F. Peters and G.~de~With, ``{Rejection-free Monte Carlo sampling for
  general potentials},'' {\em Phys. Rev. E}, vol.~85, p.~026703, 2012.

\bibitem{Faulkner2018}
M.~F. Faulkner, L.~Qin, A.~C. Maggs, and W.~Krauth, ``{All-atom computations
  with irreversible Markov chains},'' {\em J. Chem. Phys.}, vol.~149, no.~6,
  p.~064113, 2018.

\bibitem{Hoellmer2020}
P.~H\"{o}llmer, L.~Qin, M.~F. Faulkner, A.~C. Maggs, and W.~Krauth,
  ``{JeLLyFysh-Version1.0~{\textemdash} a Python application for all-atom
  event-chain Monte Carlo},'' {\em Comput. Phys. Commun.}, vol.~253, p.~107168,
  2020.

\bibitem{Erignoux2024}
C.~Erignoux, W.~Krauth, B.~Massoulié, and C.~Toninelli.
\newblock {Manuscript in preparation}.

\bibitem{Klement2019}
M.~Klement and M.~Engel, ``{Efficient equilibration of hard spheres with
  Newtonian event chains},'' {\em J. Chem. Phys.}, vol.~150, no.~17, p.~174108,
  2019.

\bibitem{Michel2020}
M.~Michel, A.~Durmus, and S.~S{\'{e}}n{\'{e}}cal, ``{Forward Event-Chain Monte
  Carlo: Fast Sampling by Randomness Control in Irreversible Markov Chains},''
  {\em J. Comput. Graph. Stat.}, vol.~29, no.~4, pp.~689--702, 2020.

\bibitem{Hoellmer2022}
P.~H\"{o}llmer, N.~Noirault, B.~Li, A.~C. Maggs, and W.~Krauth, ``{Sparse
  Hard-Disk Packings and Local Markov Chains},'' {\em Journal of Statistical
  Physics}, vol.~187, no.~3, 2022.

\bibitem{Davis1984}
M.~H.~A. Davis, ``{Piecewise-Deterministic Markov Processes: A General Class of
  Non-Diffusion Stochastic Models},'' {\em J. R. Stat. Soc. Series B Stat.
  Methodol.}, vol.~46, no.~3, pp.~353--376, 1984.

\bibitem{BierkensPDMC2017}
J.~Bierkens, A.~Bouchard-C{\^o}t{\'e}, A.~Doucet, A.~B. Duncan, P.~Fearnhead,
  T.~Lienart, G.~Roberts, and S.~J. Vollmer, ``{Piecewise Deterministic Markov
  Processes for Scalable Monte Carlo on Restricted Domains},'' {\em Statistics
  and Probability Letters}, vol.~136, p.~148{\textendash}154, 2018.

\bibitem{Bierkens2019}
J.~Bierkens, P.~Fearnhead, and G.~Roberts, ``{The Zig-Zag process and
  super-efficient sampling for Bayesian analysis of big data},'' {\em Ann.
  Stat.}, vol.~47, no.~3, pp.~1288--1320, 2019.

\bibitem{Edwards1982surface}
S.~F. Edwards and D.~R. Wilkinson, ``{The surface statistics of a granular
  aggregate},'' {\em Proc. R. Soc. A}, vol.~381, no.~1780, pp.~17--31, 1982.

\bibitem{Caputo2010}
P.~Caputo, T.~Liggett, and T.~Richthammer, ``{Proof of Aldous' spectral gap
  conjecture},'' {\em J. Amer. Math. Soc.}, vol.~23, no.~3, pp.~831--851, 2010.

\bibitem{KapferKrauth2017}
S.~C. Kapfer and W.~Krauth, ``{Irreversible Local Markov Chains with Rapid
  Convergence towards Equilibrium},'' {\em Phys. Rev. Lett.}, vol.~119,
  p.~240603, 2017.

\bibitem{Bernard2011}
E.~P. Bernard and W.~Krauth, ``{Two-Step Melting in Two Dimensions: First-Order
  Liquid-Hexatic Transition},'' {\em Phys. Rev. Lett.}, vol.~107, p.~155704,
  2011.

\bibitem{Li2022}
B.~Li, Y.~Nishikawa, P.~Höllmer, L.~Carillo, A.~C. Maggs, and W.~Krauth,
  ``{Hard-disk pressure computations—a historic perspective},'' {\em J. Chem.
  Phys.}, vol.~157, p.~234111, 12 2022.

\bibitem{Maggs2024}
A.~C. Maggs, ``{Non-reversible Monte Carlo: An example of “true”
  self-repelling motion},'' {\em Europhysics Letters}, vol.~147, no.~2,
  p.~21001, 2024.

\bibitem{Maggs2024b}
A.~C. Maggs, ``{Event-chain Monte Carlo and the true self-avoiding walk},''
  {\em arXiv:2410.08694}, 2024.

\bibitem{MaggsKrauth2022}
A.~C. Maggs and W.~Krauth, ``{Large-scale dynamics of event-chain Monte
  Carlo},'' {\em Phys. Rev. E}, vol.~105, p.~015309, 2022.

\end{thebibliography}
\end{document}